\documentclass[12pt]{article}

\usepackage{amsthm,amsmath,natbib,pgf}
\usepackage{psfrag,graphicx,verbatim,array,multicol,palatino,enumerate}
\usepackage{amsmath,amsfonts,bm,rotating,natbib,amsthm, amssymb}
\usepackage{pstricks,pst-node,setspace}
\usepackage{epsfig}
\usepackage{longtable}
\usepackage[footnotesize,bf]{caption}

\begin{document}

\begin{center}

\begin{Large}

{\bf Refining the Protein-Protein Interactome Using Gene Expression Data}

\end{Large}

\vspace{1.5cm}

Sira Sriswasdi\footnote{Genomics and Computational Biology Graduate Group, University of Pennsylvania \\ {\tt{sirasris@mail.med.upenn.edu}}} and Shane T. Jensen\footnote{Department of Statistics, The Wharton School, University of Pennsylvania \\ {\tt{stjensen@wharton.upenn.edu}}} \\

\vspace{0.25cm}

\end{center}

\vspace{1cm}

\begin{abstract}

Proteins interact with other proteins within biological pathways, forming connected subgraphs in the protein-protein interactome (PPI).  Proteins are often involved in multiple biological pathways which complicates interpretation of interactions between proteins. Gene expression data can assist our inference since genes within a particular pathway tend to have more correlated expression patterns than genes from distinct pathways. We provide an algorithm that uses gene expression information to remove inter-pathway protein-protein interactions, thereby simplifying the structure of the protein-protein interactome. This refined topology permits easier interpretation and greater biological coherence of multiple biological pathways simultaneously.  
\end{abstract}

\doublespacing

\section{Introduction}

The protein-protein interactome (PPI) is a large graph where proteins are nodes and edges between these nodes represent all known interactions between proteins.   In cases where proteins interact in order to drive a particular biological process, the connected nodes of a PPI can represent an entire biological pathway.   However, inferring a biological pathway from the PPI is complicated by the fact that many proteins are involved in multiple biological functions.    Thus, a connected subgraph of the PPI must be viewed as a mixture of smaller graphs that each represent a particular pathway.   It is the goal of this paper to refine the PPI by isolating these smaller graph components which are more likely to contain just a single pathway. 

Our primary tool for this endeavor is gene expression data, which allows us to identify pairs of genes with highly correlated expression patterns.  In general, gene pairs are more likely to have correlated expression if they belong to the same biological pathway, which gives us a mechanism for refining the PPI to isolate individual pathways.   We introduce a procedure for reducing large connected components of the PPI into smaller groups with higher connectivity that represent a single pathway.

\section{Methods}\label{methods}

\begin{figure}[h]
  \centerline{\includegraphics[height=4in]{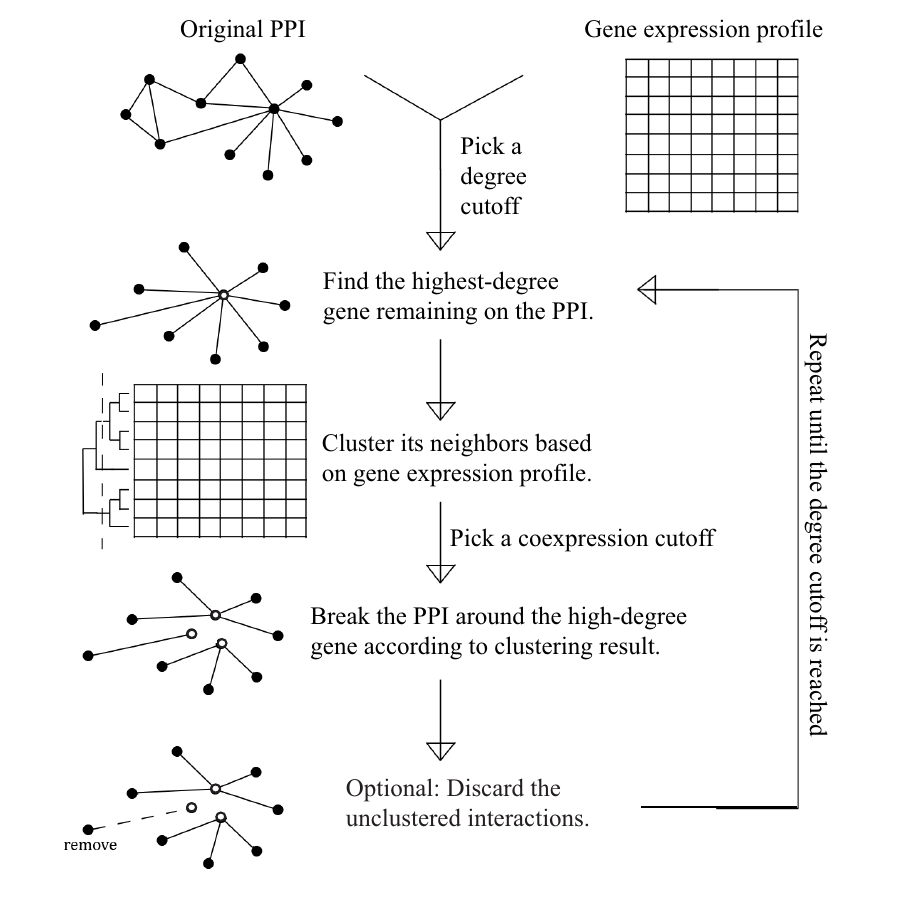}}
 \caption{Overall framework of our PPI refining procedure.\label{fig:framework}}
\end{figure}

The input for our procedure is a protein-protein interactome and a set of gene expression profiles.  Our data sources are given in Appendix~\ref{datasources}.   The overall framework of our algorithm is shown in Figure~\ref{fig:framework}.

Our algorithm focuses initially on proteins with the highest degree in the PPI as potential multi-pathway proteins.  Let $A$ be the protein that currently has the largest number of connections in the PPI.  We denote $\mathcal{N}(A)$ as a set containing all proteins connected to $A$ via protein-protein interactions.    

For each pair of proteins $i$ and $j$ in $\mathcal{N}(A)$, we calculate the correlation $\rho_{ij}$ of their gene expression patterns.   An agglomerative hierarchical clustering of the proteins in $\mathcal{N}(A)$ is performed using $d_{ij} = 1 - |\rho_{ij}|$ as the distance metric.   A pre-specified threshold $\theta_\text{cor}$ is used to convert this hierarchical clustering into a partition of disjoint subsets $N_1,\ N_2,\ \dots,\ N_m$ of highly correlated proteins, as well as an extra subset $N_{m+1}$ containing unclustered proteins.   

We now proceed under the assumption that each disjoint subset $N_1,\ N_2,\ \dots,\ N_m$ of proteins with highly correlated gene expression is a group of proteins in $\mathcal{N}(A)$ belonging to the same pathway.    We remove inter-pathway connections within $\mathcal{N}(A)$ by replacing protein $A$ in the PPI with duplicates $A_1,\ A_2,\ \dots,\ A_m$, where $A_i$ retains only the connections between $A$ and proteins in $N_i$.    We also discard all connections between protein $A$ and proteins contained in the unclustered set $N_{m+1}$.  

These expression clustering and network reduction steps are repeated for all highly-connected proteins in the protein-protein interactome.   We terminate the algorithm when no protein in the refined PPI contains more connections than a pre-specified degree cutoff $\theta_\text{deg}$.    

In the following section, we examine the performance of our method for refining the PPI under different settings for our two user-specified parameters: the cutoff $\theta_\text{deg}$ for the largest degree protein in the refined PPI and the threshold $\theta_\text{cor}$ for partitioning highly correlated gene expression profiles into disjoint clusters.

\section{Results}

We examined four different versions of our refined PPI corresponding to two settings of the largest degree protein cutoff  $\theta_\text{deg} \in \{4,9\}$ and two settings of the co-expression cutoff $\theta_\text{cor} \in \{0.4,0.6\}$.   For each combination of these $\theta_\text{deg}$ and $\theta_\text{cor}$, we produced a refined PPI using the method in Section~\ref{methods}.  We compare these refined PPIs to the original PPI in terms of both KEGG pathways and the GO ontology. 

\subsection{Evaluation using KEGG pathways}
We use the KEGG pathway database \cite[]{KEGG1,KEGG2} to evaluate the performance of our method. The KEGG database contains high-confidence manually-reviewed mapping information between genes and biological pathways. 

Out of the 2940 common genes between our protein-protein interaction network and our gene expression datasets, 1033 genes can be mapped to 90 distinct pathways in the KEGG dataset.  For our evaluation, we focus on the pathways which contain at least 10 protein-protein interactions, which gives us 26 pathways covering 761 genes. 

For each KEGG pathway, we define a {\it connectivity score} as the proportion of gene pairs belonging to that KEGG pathway that are also connected in the PPI.    We then calculate, for each KEGG pathway $i$, the {\it enrichment} of our refined PPI, 
\begin{eqnarray*}
\text{enrichment}_i = \frac{\text{connectivity score of pathway} \, i \, \text{in refined PPI}}{\text{connectivity score of pathway} \, i \,\text{in original PPI}}
\end{eqnarray*}

\begin{figure}[h]
  \centerline{\includegraphics[height=3in]{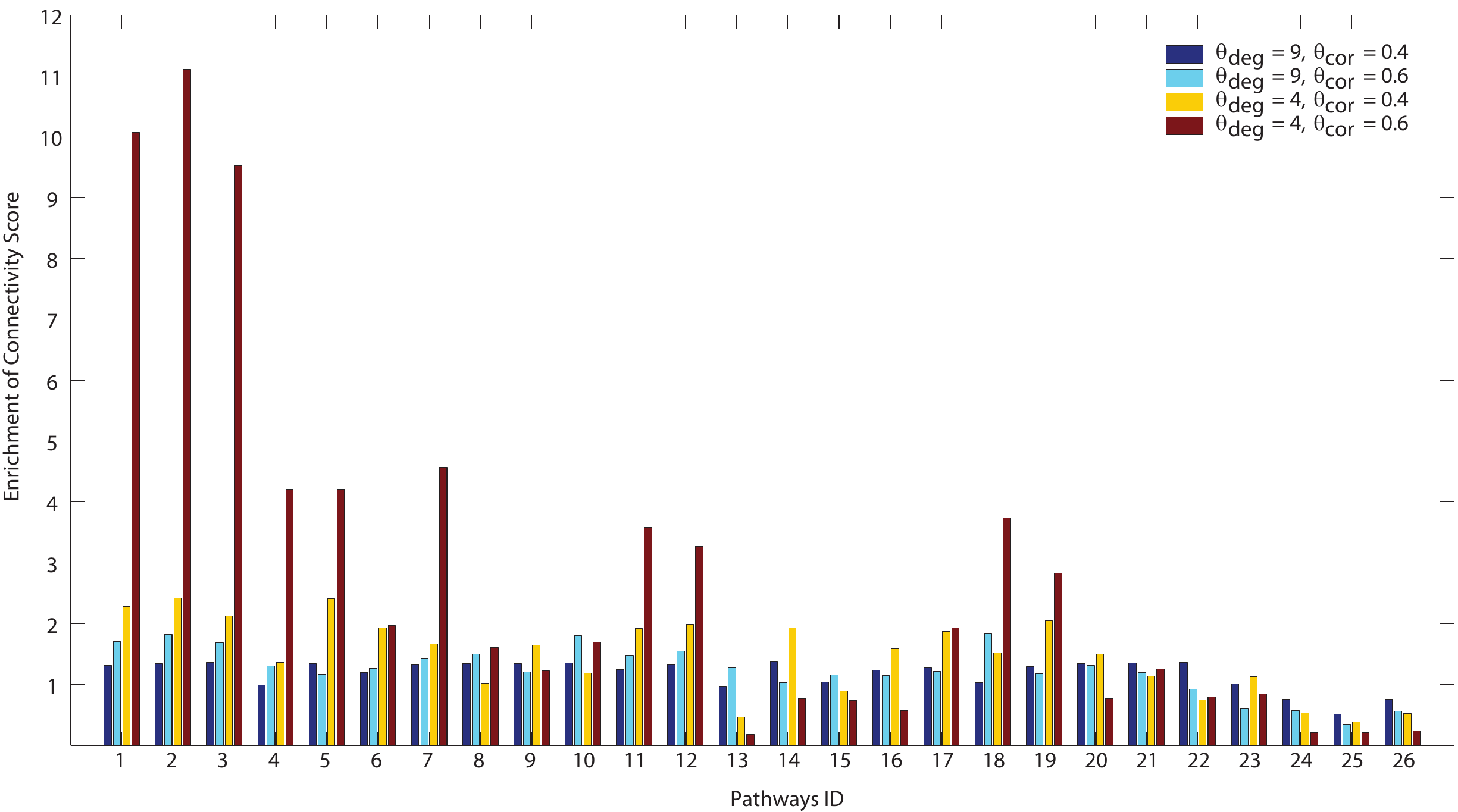}}
  \caption{Enrichment of the connectivity scores from our refined PPIs to the original PPI.  Different colors correspond to different settings of input parameters $\theta_\text{deg} \in \{4,9\}$ and $\theta_\text{cor} \in \{0.4,0.6\}$.  \label{DNplot}}
\end{figure}

We compare the enrichment between our four different refined PPIs to the original PPI for all 26 KEGG pathways in Figure~\ref{DNplot}.   In this figure, we see substantial differences between four different settings of our input parameters.  For example, the red bars ($\theta_\text{cor} = 0.6$ and $\theta_\text{deg}$ = 4) show very high enrichment for many of the KEGG pathways (e.g. pathways 1-5 and 7) but also low enrichment relative to the other settings for other KEGG pathways (e.g. pathways 20 and 24-26).    

Despite the differences between our refined PPIs, the primary observation from Figure~\ref{DNplot} is that the enrichment exceeds one for each of our parameter settings in most KEGG pathways.  In other words, our refined PPIs are enriched relative  to the original PPI in most KEGG pathways, regardless of our choice of input parameters.

To quantify the statistical significance of the enrichment in Figure~\ref{DNplot}, we employ a randomized control version of our algorithm.  In this randomized control version, genes are randomly permuted after expression clustering to randomize the clustering result while preserving the number of genes in each cluster.    

Our refined PPI shows significant (at the 5\% level) enrichment over the original PPI relative to these randomized controls in many of these KEGG pathways.  For example, under the parameter setting of $\theta_\text{cor} = 0.6$ and $\theta_\text{deg}$ = 4 (red bars in Figure~\ref{DNplot}), there is significant enrichment at the 5\% level in thirteen pathways (pathways 1-5, 7-8, 10-12 and 17-19). 

The performance of our method varies substantially between KEGG pathways, as evidenced by both Figure~\ref{DNplot} and the fact that only 13 out of 26 pathways were significantly enriched in our refined PPI.  Much of this variation can be explained by the central assumption of our method that only genes with highly correlated expression patterns should be connected in the PPI.   

\begin{figure}[h]
  \centerline{\includegraphics[height=2.5in]{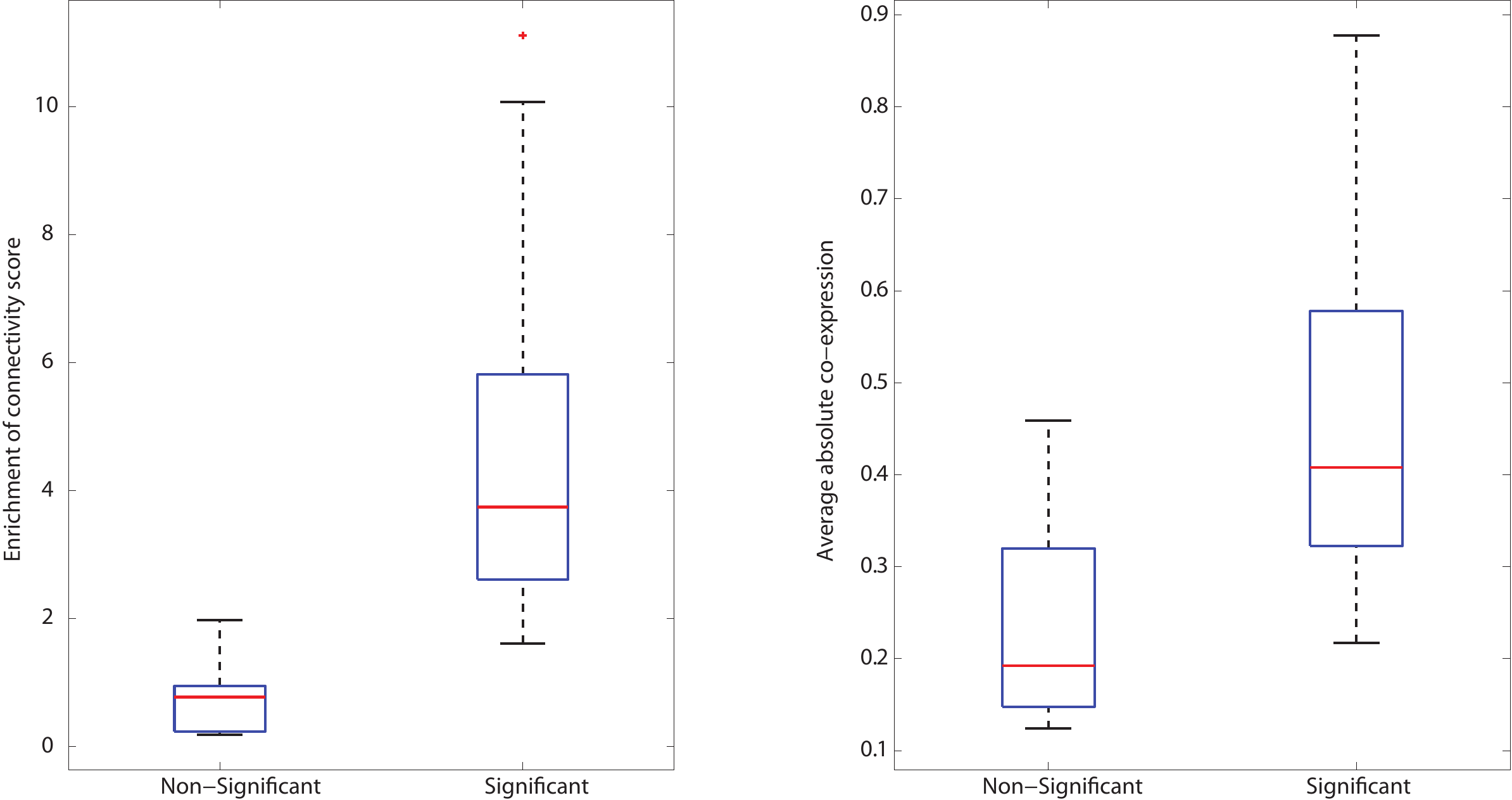}}
  \caption{Boxplots comparing the (left) connectivity score enrichment and (right) average absolute co-expression between pathways that were significantly enriched versus pathways that were not significantly enriched\label{twoclass}}
\end{figure}

In Figure~\ref{twoclass}, we compare the KEGG pathways that were significantly enriched in our refined PPI to the pathways that were not significantly enriched, using the refined PPI from setting $\theta_\text{cor} = 0.6$ and $\theta_\text{deg}$ = 4.   In the left panel, we show that the enrichment scores are much higher in the pathways that were significantly enriched relative to our randomized control algorithm.  

In the right panel, we see that average absolute co-expression is much higher between genes in the significantly enriched KEGG pathways, which confirms that our algorithm performs better in pathways where connected proteins also have highly correlated expression patterns.  

In summary, our method for refining the PPI leads to higher connectivity in most KEGG pathways we examined, and significantly higher connectivity (relative to a randomized control algorithm) in half of the KEGG pathways.   Pathways that were not significantly enriched by our method tended to be pathways containing proteins that had less correlated gene expression patterns.

\subsection{Evaluation using gene ontology}

We also examine our method for refining the PPI by examining the gene ontology (GO) database \cite[]{Go00}, which is a multi-level collection of biological terms assigned to specific genes.   The GO database contains three types of biological terms: cellular component, molecular function, and biological process.    Molecular function is the most specific type, but many proteins either lack molecular function annotations or do not share a common annotation with other proteins.   In contrast, most proteins are annotated with a cellular component GO term, but this feature is too broad to be particularly informative.   We focus our analysis on {\it biological process} GO terms as the GO type most closely related to our goal of isolating biological pathways.  

For a group of proteins in the protein-protein interactome, we define an evaluation metric called the {\it deepest common GO depth}.   The {\it deepest common GO depth} is the depth in the GO hierarchy of the deepest GO term that is common to all proteins in a connected group.   The GO hierarchy becomes more specific as the depth increases, so large common GO depths are indicative of a group of proteins that have high coherence in terms of their biological processes.   In contrast, a GO depth of zero indicates that a group of proteins have such low coherence that one most go all the way to the root node of the GO hierarchy to find a common GO term for that group of proteins. 

\begin{figure}[h]

  \centerline{\includegraphics[height=3.25in]{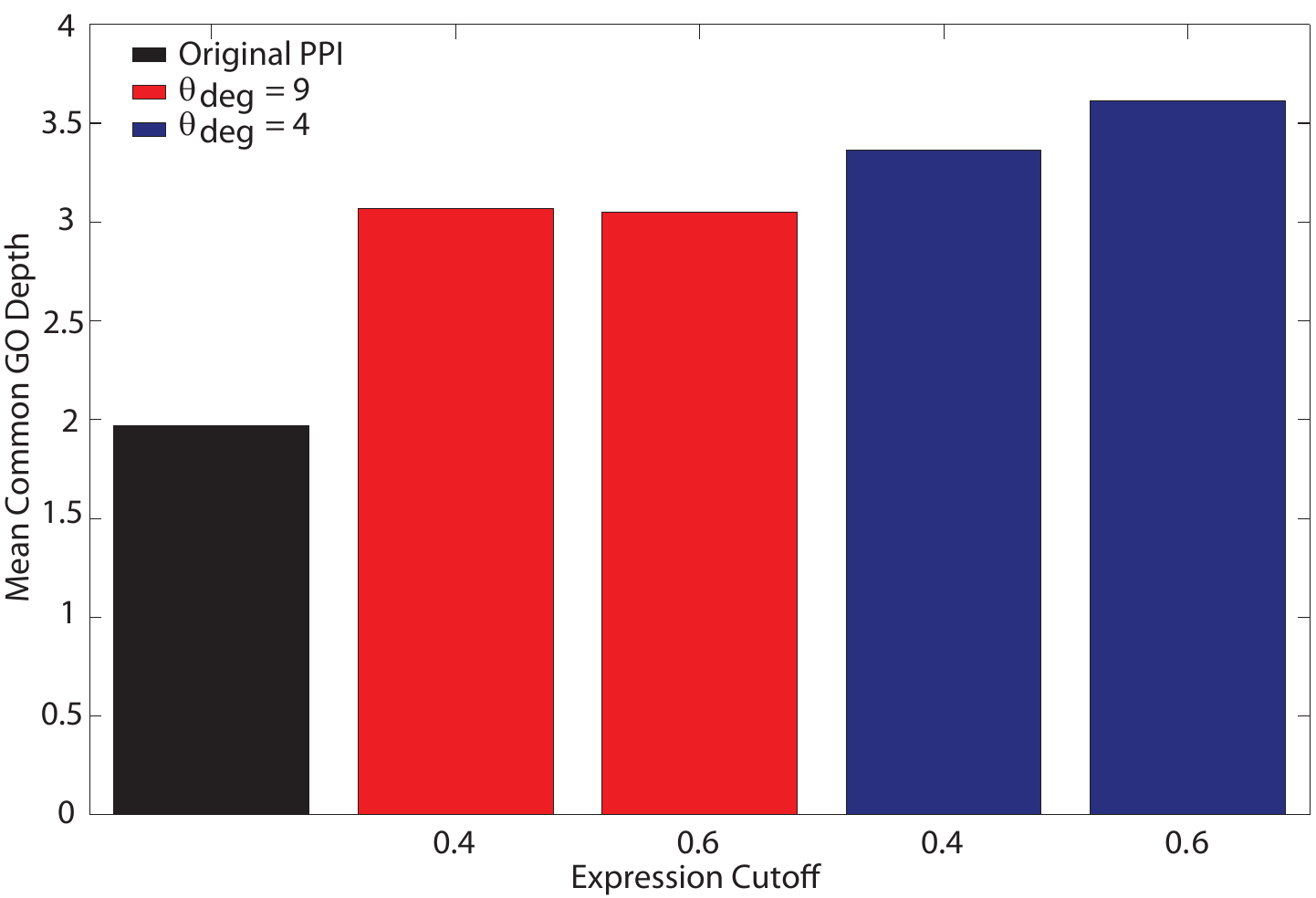}}
    
  \caption{Comparing average deepest common GO depth between the original PPI (black) and different settings of our input parameters\label{GOcomp}.  Larger values correspond to greater coherence with the gene ontology.}
\end{figure}

For each protein in a particular PPI, we compute the  deepest common GO depth for the group of proteins consisting of that protein and its direct interacting neighbors.   We then average those deepest common GO depths over all proteins in that PPI.     In Figure~\ref{GOcomp}, we compare the {\it average deepest common GO depth} of the original PPI  to several refined PPIs corresponding to different choices of the absolute co-expression cutoff  $\theta_\text{cor}$ parameter and the degree cutoff $\theta_\text{deg}$ parameter. 

We see that for all choices of input parameters $\theta_\text{cor}$ and $\theta_\text{deg}$, the refined PPI from our procedure shows a dramatically larger {\it average deepest common GO depth}.    This result suggests that the refined protein connections from our procedure have a much greater coherence in their biological processes compared to the original PPI.     

\section{Discussion}

We have presented a method for refining the protein-protein interactome using gene expression data.   We can produce refined PPIs from our procedure under many different choices of two input parameters, the cutoff $\theta_\text{deg}$ for the largest degree protein in the refined PPI and the threshold $\theta_\text{cor}$ for partitioning highly correlated gene expression profiles into disjoint clusters.     For each combination of parameter settings we examined, our refined PPI shows greater GO ontology coherence (Section 3.2) and higher enrichment of connectivity in most KEGG pathways (Section 3.1).
 
\begin{figure}[h]

  \centerline{\includegraphics[height=3.25in]{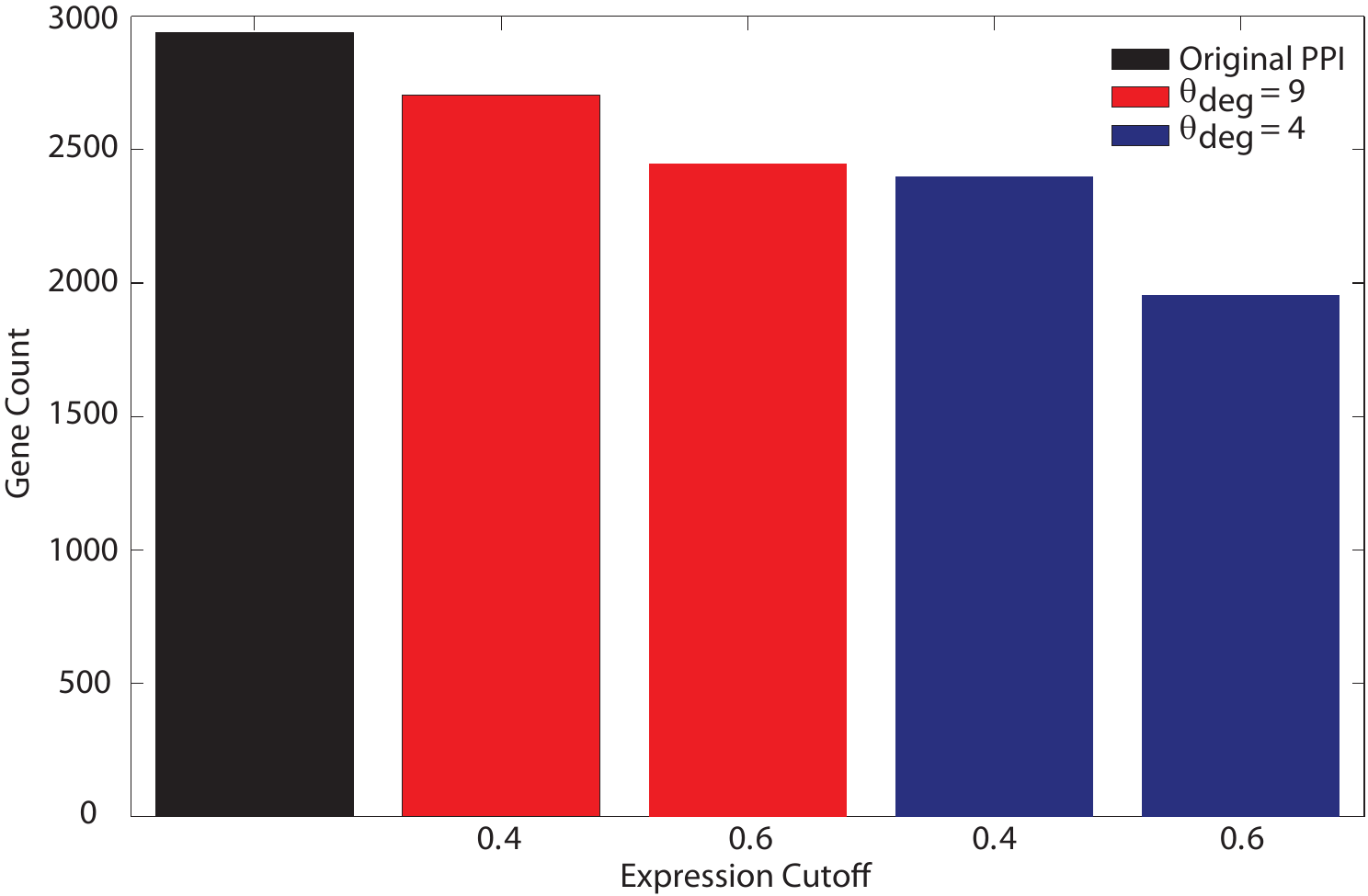}}

  \caption{Number of proteins contained in the refined PPI using different parameter settings \label{genecount}}
\end{figure}

Although our procedure results in a PPI with greater biological coherence, there is one sacrifice: some proteins are removed completely from the refined PPI due to all of their connections to other proteins being removed.    We must balance our increase in biological coherence with the reduction in the number of proteins contained in the PPI.    In Figure~\ref{genecount}, we give the number of proteins in the original PPI as well as the number of proteins remaining in the PPI for each parameter setting examined in Section 3.

Not surprisingly, stricter choices of the threshold parameters $\theta_\text{cor}$ and $\theta_\text{deg}$ lead to a PPI that has many proteins removed.   Based on these results, we suggest parameter values of $\theta_\text{cor} = 0.8$ and $\theta_\text{deg} = 4$ as a good compromise that gives increased biological coherence without the removal of too many proteins from the PPI.    We provide our refined PPI under these parameter settings along with code for producing refined PPIs under other parameter settings will be made available upon publication.

\begin{appendix}

\section{Data sources}\label{datasources}

Our protein-protein interactome data is a combination of two datasets, from  \cite{KroCagYu06short} and \cite{GavBosKra02short}.   The data has since been moved to the "Protein complex in yeast" website:

http://yeast-complexes.embl.de/tmp/socio-affinities.dat.gz

The gene expression data used for our analysis is from \cite{CheJenSto07} and can be downloaded at:

http://www-stat.wharton.upenn.edu/$\sim$stjensen/research/cogrim/expression.genes.txt

\end{appendix}

\bibliography{references}

\begin{thebibliography}{}

\bibitem[Chen \emph{et~al.}(2007)Chen, Jensen, and Stoeckert]{CheJenSto07}
Chen, G., Jensen, S.~T., and Stoeckert, C.~J. (2007).
\newblock Clustering of genes into regulons using integrated modeling -
  {COGRIM}.
\newblock \emph{Genome Biology} \textbf{8}, R4.

\bibitem[Gavin \emph{et~al.}(2002)]{GavBosKra02short}
Gavin, A.-C. \emph{et~al.} (2002).
\newblock Functional organization of the yeast proteome by systematic analysis
  of protein complexes.
\newblock \emph{Nature} \textbf{415}, 6868, 141--147.

\bibitem[Kanehisa and Goto(2000)]{KEGG1}
Kanehisa, M. and Goto, S. (2000).
\newblock {KEGG}: Kyoto encyclopedia of genes and genomes.
\newblock \emph{Nucleic Acids Research} \textbf{28}, 27--30.
\newblock {\bf Downloaded: February, 2012}.

\bibitem[Kanehisa \emph{et~al.}(2012)Kanehisa, Goto, Sato, Furumichi, and
  Tanabe]{KEGG2}
Kanehisa, M., Goto, S., Sato, Y., Furumichi, M., and Tanabe, M. (2012).
\newblock {KEGG} for integration and interpretation of large-scale molecular
  datasets.
\newblock \emph{Nucleic Acids Research} \textbf{40}, D109--D114.

\bibitem[Krogan \emph{et~al.}(2006)]{KroCagYu06short}
Krogan, N. \emph{et~al.} (2006).
\newblock Global landscape of protein complexes in the yeast saccharomyces
  cerevisiae.
\newblock \emph{Nature} \textbf{440}, 7084, 637--643.

\bibitem[{The Gene Ontology Consortium}(2000)]{Go00}
{The Gene Ontology Consortium} (2000).
\newblock Gene ontology: tool for the unification of biology.
\newblock \emph{Nature Genetics} \textbf{25}, 25--29.
\newblock {\bf Downloaded: December, 2009}.

\end{thebibliography}

\end{document}